\documentclass[11pt]{article}
\usepackage{graphicx}
\usepackage[margin=1.25in]{geometry}
\usepackage[usenames,dvipsnames]{color}
\usepackage{url}
\usepackage[colorlinks = true,
            linkcolor = blue,
            urlcolor  = blue,
            citecolor = blue,
            anchorcolor = blue]{hyperref}
\usepackage[utf8]{inputenc}
\usepackage{dcolumn}   
\usepackage{bm}        
\usepackage{amssymb}   
\usepackage{amsmath}
\usepackage[caption=false]{subfig}
\usepackage[nomargin,inline,marginclue,draft]{fixme}

\def\Title#1{\begin{center} {\LARGE #1 } \end{center}}
\def\Author#1{\begin{center}{ \sc #1} \end{center}}
\def\Address#1{\begin{center}{ \it #1} \end{center}}

\newenvironment{Abstract}{\begin{quotation} \begin{center}
                       ABSTRACT
     \end{center}\bigskip  }{\end{quotation}}

\begin{document}


\Title{Data Storage for HEP Experiments \break 
in the Era of High-Performance Computing}

\smallskip 

\Author{Amit Bashyal and Peter van Gemmeren}
\Address{ Argonne National Laboratory }
\smallskip
\Author{Saba Sehrish and Kyle Knoepfel}
\Address{ Fermi National Accelerator Laboratory }
\smallskip
\Author{Suren Byna and Qiao Kang}
\Address{ Lawrence Berkeley National Laboratory }

\Author{On Behalf of the HEP-CCE IOS Group}

\begin{Abstract}
\noindent As particle physics experiments push their limits on both the energy and the intensity frontiers, the amount and complexity of the produced data are also expected to increase accordingly. With such large data volumes, next-generation efforts like the HL-LHC and DUNE will rely even more on both high-throughput (HTC) and high-performance (HPC) computing clusters.   Full utilization of HPC resources requires scalable and efficient data-handling and I/O. For the last few decades, ROOT has been used by most HEP experiments to store data. However, other storage technologies like HDF5 may perform better in HPC environments. Initial explorations with HDF5 have begun using ATLAS, CMS and DUNE data; the DUNE experiment has also adopted HDF5 for its data-acquisition system. This paper presents the future outlook of the HEP computing and the role of HPC, and a summary of ongoing and future works to use HDF5 as a possible data storage technology for the HEP experiments to use in HPC environments. 
\end{Abstract}


\def\thefootnote{\fnsymbol{footnote}}
\setcounter{footnote}{0}


\section{Introduction}

With the growing computational demands of the upcoming HEP experiments, the role of high-performance computing (HPC) will also increase.  HPC systems, however, are different than the traditional grid-computing system with highly customized architectures and hardware. To use HPC systems to meet the HEP community's computational demands, tools and technologies must be adopted to utilize HPC resources. The organizational complexity of HEP data products and workflows, however, makes complete adoption of such technologies non-trivial.  We thus outline an approach that motivates the use of HPC systems for processing contexts where significant computational capabilities are required, such as data reconstruction.  However, for further-downstream contexts such as data analysis, we support HEP experiments in their continued use of technologies such as ROOT~\cite{ROOTCERN}, which has been a workhorse of this community for the last two decades. 

This white paper discusses the role of HPC in light of the growing computational needs of HEP experiments in section \ref{HEP computing context} . In section~\ref{HDF5 in HEP}, we discuss our exploratory work of using HDF5, one of the HPC-friendly I/O libraries, to read/write HEP experiment data. In section \ref{Future Directions}, we discuss some of the avenues that can be explored by the HEP community to introduce and eventually use HPC systems in their workflows.  


\section{HEP Computing Context}
\label{HEP computing context}

The requirement to handle big data has pushed the HEP field to explore and innovate data-computing and storage technology. The Open Science Grid \cite{osg07} and the Worldwide LHC Computing Grid (WLCG) \cite{Bird:2011zz} are examples of how the HEP community (notably through Fermilab and CERN) is firmly committed to seeking solutions to meets its computational and storage demands.  These endeavors, which utilize high-throughput computing (HTC), have been critical ingredients in meeting experiments' physics goals such as discovering the Higgs boson.  

\begin{figure}[htb]
    \centering
    \includegraphics[width=0.8\linewidth]{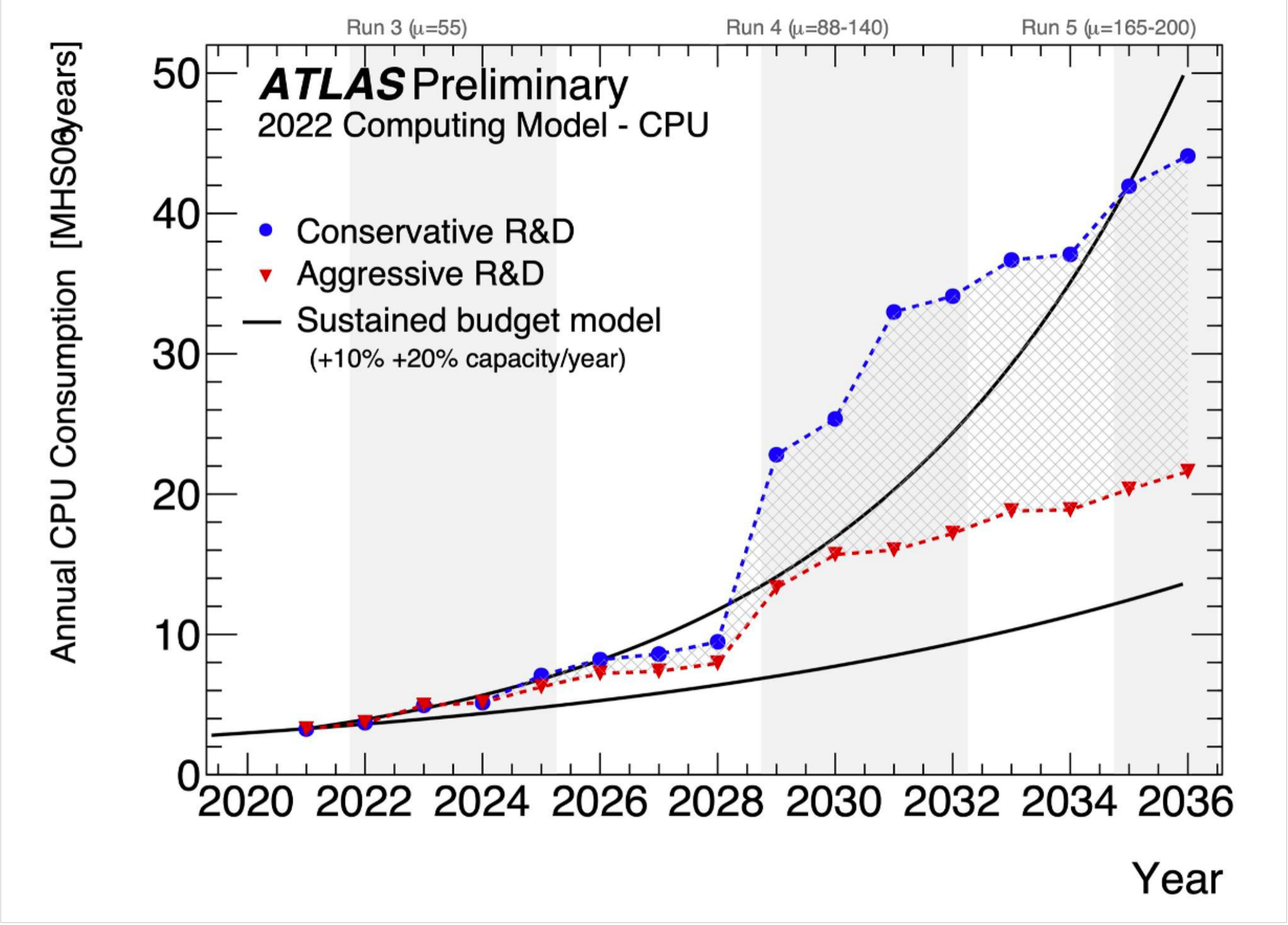}
    \caption{Projected CPU consumption rate from 2020 to 2036 for the ATLAS experiment. The area between the blue and the red dot line shows the range of CPU consumption estimation based on conservative (blue) to aggressive (red) R\&D in the computing infrastructure. Conservative scenario assumes the current person power maintained at current level with the current level of expertise. Aggressive R\&D assumes increase in expertise either by allowing current person power to increase the time in software R\&D or bringing new people.  The top (bottom) black line shows the impact of 10\% (20\%) increase in budget annually in the R\&D. Figure taken from~\cite{Marshall:2800627}. }
    \label{fig:ATLAS}
\end{figure}

Up until Run II of the LHC, the WLCG system met data-processing needs at a rate that was commensurate with those of data collection. By the end of Run 5, however, the ATLAS and CMS experiments estimate that the collected data be 10 times larger than was collected during the first 3 runs~\cite{Marshall:2800627}. Projections made by ATLAS~\cite{Marshall:2800627} and CMS~\cite{Oliver022} show that without a radical shift in the data-management and -processing approach, the current computing infrastructure will not be able to accommodate the storage and computing needs of these experiments in future LHC runs.  To illustrate, Figure~\ref{fig:ATLAS} shows the projected annual CPU consumption of the ATLAS experiment from 2020 to 2036. As can be seen, with only modest improvements in the computing and software infrastructure, the year-by-year CPU consumption rate will increase such that the current infrastructure will not be able to sustain even the most optimistic scenario. To comfortably accommodate the goals of future LHC runs, a somewhat aggressive R\&D program is thus required.

\begin{figure}
    \centering
    \includegraphics[width=0.8\linewidth]{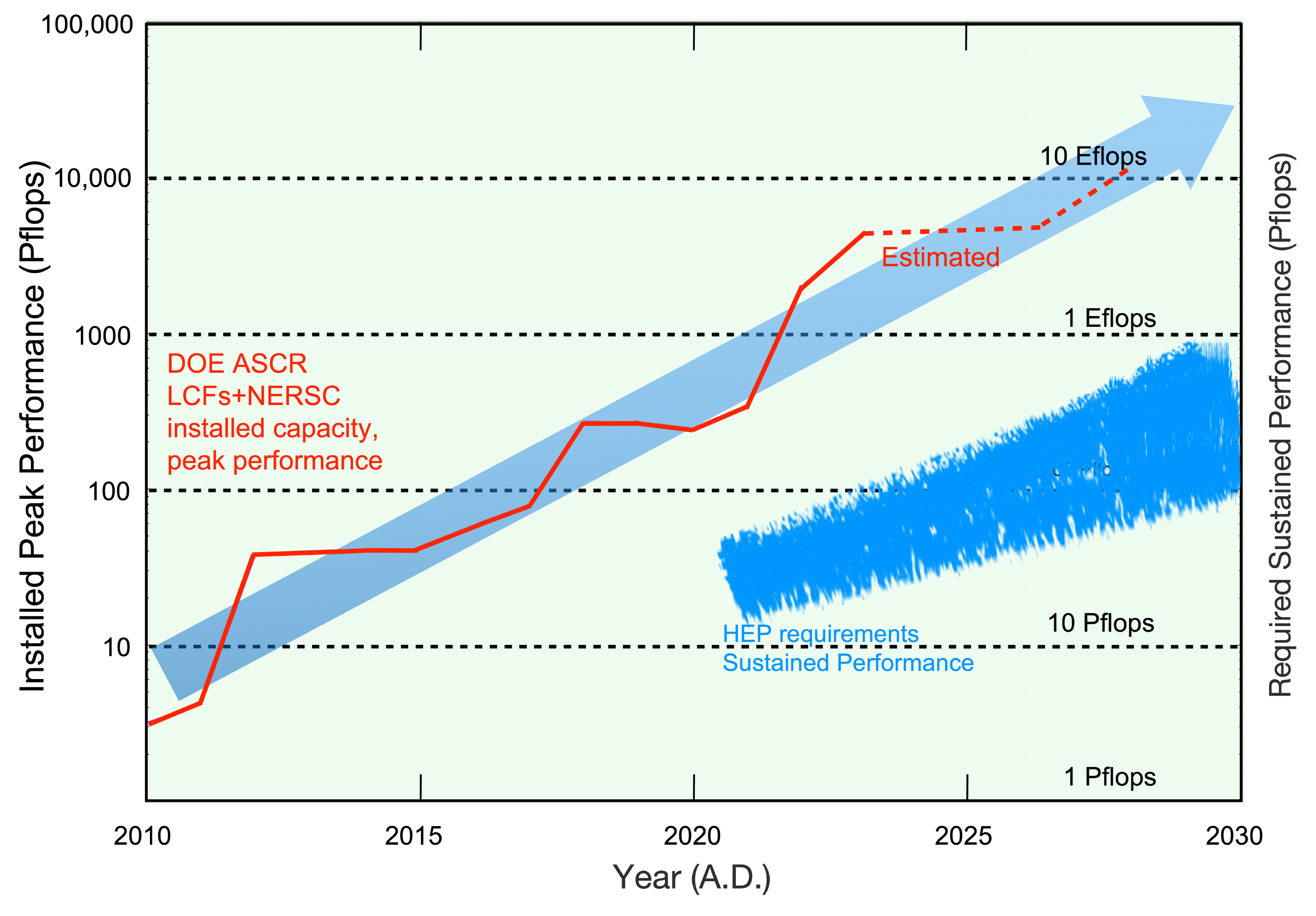}
    \caption{Projected peak CPU performance of the various DOE super computers (red line) and the estimated HEP CPU  resource requirements assuming sustained performance (blue shade). Figure taken from~\cite{HEPCCE022}.}
    \label{fig:HEPandHPC}
\end{figure}

Intensity-frontier experiments have slightly different computing challenges compared to those of the LHC.  Although the DUNE experiment will be larger in physical size than the LHC experiments (with up to four 10 kT active volumes of liquid argon), DUNE's far detector \cite{DUNE:2020lwj} will have fewer channels than those of the LHC experiments. The data rates from DUNE's channels, however, will be significant--anywhere from a few GB for beam-induced readouts to over 100 TB for supernova readouts.  DUNE is thus a good candidate for utilizing the parallel and accelerated data-processing capabilities of HPC resources~\cite{Laycock021}.  Furthermore, as the US Department of Energy (DOE) is investing heavily in HPC projects among its national labs, the HEP community should avail itself of the HPC resources provided by the DOE.

The HPC ecosystem promises an efficient data-management and -computation environment that could greatly benefit the HEP community.  As shown in Figure~\ref{fig:HEPandHPC}, HPC resources could meet the HEP community's CPU requirements through the DUNE and LHC Run 5 eras. The implication, however, is that HEP experiments would need to invest in the computing R\&D efforts necessary for using HPC systems.  The HEP community would need to understand not only how to adjust its current software for HPC systems, but also how that software would be deployed, how to use the parallel I/O and parallel computing features of the HPC machine, and what data storage mechanisms are best used in an HPC environment.

\subsection{High Performance I/O and Storage Requirements}
 
HPC systems use parallel file systems that enable efficient data storage and access. High-level parallel I/O libraries, such as HDF5~\cite{hdf5,Folk:2011:OHT, Byna:2020:ExaHDF5:JCST} and PnetCDF~\cite{pnetcdf:2003}, provide interfaces for applications to use parallel file systems without exposing parallelism details to the user. In addition to the already complex parallel I/O, there are several architectural changes expected in upcoming HPC systems. Traditional storage sub-systems have been based on hard-disk drives (HDD), which are much slower than the memory hardware of compute nodes. The latency of HDD-based file systems is significantly high as a result. To mitigate the impact of this latency on application performance, the latest HPC systems are adding faster layers of storage (e.g.,\ non-volatile memory, solid-state drives) between a compute node's memory and traditional parallel file systems. These additional layers of storage and their file systems further increase the complexity of using I/O on modern HPC systems. 
 
Utilization of the HPC computing resources requires storing and accessing these new types of storage systems efficiently. Below is a list of some HEP-specific requirements for using HPC storage systems. 

\begin{itemize}
    \item \emph{Parallelism:} Stored data should be able to be accessed by multiple parallel processes. This requires multiple parallel processes to be able to do I/O operations (e.g.\ opening a file, reading and writing buffers) on the data concurrently.
    \item \emph{Data model:} HEP data products are often complex C++ objects that cannot be serialized easily (currently we can depend on ROOT to serialize the data-products.) Either the storage software should provide tools to serialize the complex data products (highly unlikely) or the data products should be modeled to be friendly with the storage systems.
    \item \emph{Visibility:} HPC-friendly data storage systems provide high level libraries to fully utilize the HPC resources. 
    However, users should be able to customize and configure the parameters depending on the requirement of the HEP experiment.
    \item \emph{Usability:} The HEP community is accustomed to ROOT's user-friendliness, reasonably thorough code documentation, and robust user forums to share issues, solutions, and learn of new features.   The community will expect similar attention to usability with any other storage format chosen for HPC environments.
\end{itemize}

This is not an exhaustive list of requirements and the HEP community may discover further requirements as the role of HPC increases in HEP workflows. 

\section{Using HDF5 Storage for HEP Data}\label{HDF5 in HEP}

To explore using HPC I/O libraries to store HEP data, we begin with the HDF5 library~\cite{hdf5}.  We work with existing HEP data models and frameworks to design an approach that is experiment-agnostic.  We also make basic assumptions regarding the layout of experiment data:

\begin{itemize}
    \item The fundamental unit for processing HEP data is the \emph{event}, which contains \emph{data products}---constructs that represent physically meaningful information such as energy deposits (or hits), trigger times, and information related to incident beams.
    \item Events can be grouped into \emph{subruns} or \emph{luminosity blocks}, which in turn are grouped into \emph{runs}.
\end{itemize}

The representations of data products can be complex, which can be non-trivial to map to a storage backend.  Like other technologies, HDF5 does not provide the near-automatic C++ type support the HEP community has relied upon through ROOT over the last few decades.  The following discusses our efforts in mapping HEP data to HDF5.

\subsection{Mapping HEP Data to HDF5}

For a baseline result, we map HEP data to HDF5's serial I/O capabilities.  HDF5 provides the hierarchical data-model just like the directories, sub-directories and files. This model is similar to that provided by the ROOT framework, which groups stored data according to {\tt TBranch}s, {\tt TTree}s, and {\tt TFile}s.  We are exploring two approaches for the HDF5-based design: one that is data-product based and event based.

The data-product based approach is similar to how data is represented as  {\tt TBranch} objects in a ROOT {\tt TTree}. For our use case, the input data for HDF5 is considered to be already serialized using ROOT serialization. 

Each data-product is assigned a separate HDF5 dataset. This is similar to writing different data-products in different {\tt TBranch}s. Data can be aggregated in batches before writing them into HDF5 datasets or written one at a time.
 
Event boundaries are identified by storing the size of each data-product in a different HDF5 dataset. 

This method requires $2N$ HDF5 datasets for $N$ data-products and at least one additional HDF5 dataset to store the sizes of the data-products.

The event-based approach stores one complete event as a serialized byte stream and product boundaries are stored in a different dataset. Hence this method requires only three HDF5 datasets to store the data with all the relevant information.\\
Data product based approach has better compression and allows access to individual data-product without having to read the whole event. Typically HEP experimental root files have large number of data-products. Since the data-products are smaller in sizes, they need to be aggregated over multiple events which requires more memory footprint.\\
Event based method requires only two HDF5 data-sets for each event. Compared to the data-product based method, this method requires only two I/O process to read/write each event. However, this also means that the users will have to read the whole event to access necessary data-products.\\

Both event based and data-product based methods have pros and cons and we will improve these methods and explore additional optimization strategies in the upcoming days.



\subsection{Proof of Concept}
We have developed a multi-threaded test framework based on CMS's framework as part of this on-going work \cite{ROOT-Serialization}. This framework supports any number of storage formats; and can read from any supported format and write to any supported format. Any new format can be easily added by providing the necessary base-class interface implementation.  
This framework supports the study of scaling of the I/O performance with different formats on different systems, memory usage, file size and compression in both HPC and HPC environment. To study the performance of HDF5 with HEP data, we have added event-based and data-product based modules. \\
This frameowrk supports the data from ATLAS, CMS and the DUNE experiments so far.

However, further performance studies and the comparison of HDF5 with other other formats (including ROOT) are needed in the future. Furthermore, additional exploration is needed in the areas like optimization of tuning parameters and data layouts for HEP data, parallel I/O and compression in HDF5.

\subsection{Leveraging Parallel HDF5}

HDF5 provides supports for the MPI \cite{MPICH022} based parallel I/O which allows users to effectively utilize the HPC resources. Users can utilize the collective I/O libraries supported by HDF5 which allows multiple processes to write in a single file saving additional computational resources needed to merge several small files into few larger ones as done traditionally in the HEP field. HEP frameworks are usually multi-threaded and leveraging on parallel HDF5 will require further research and studies. 

\section{Future Directions}
\label{Future Directions}

Over the last 2 decades, ROOT has been the work horse of the HEP experiments in terms of data processing, storage and analysis. The tools developed by ROOT has made the work flow of the HEP experiments much smoother and easier. Until now, the data processing and storage of the HEP experiments overwhelmingly depend upon the HTC systems in which ROOT performs best. However, as discussed in this paper, the computational needs of the HEP fields over the next decade can only be fulfilled by utilizing HPC systems which have much different storage hardware and architectural design where other I/O libraries and file format like HDF5 can perform better. HEP experiments need to identify the areas where these HPC-friendly storage systems can be adopted directly or with minimal work. \\
 One of the challenges of the HEP experiment data is the serialization of data-products. HEP data- products are often complex C++ objects that are hard to couple directly with the HDF5 or other HPC-friendly I/O libraries. Currently, we rely on ROOT to serialize these complex data-products which can then be saved in the HDF5 format for our studies. However, we should identify the areas where the data-products are simple enough to be directly stored in these file formats. For example, DUNE directly stores its raw data into the HDF5 format.
 In the long run, HEP experiments should map their data-models to the HPC-friendly file formats for the computationally intensive processes like full detector simulation where parallel I/O performance is crucial. \\

\section{Summary}

HPC systems will be able to fulfill the HEP computational needs long into the future and the HEP community will see the increased role of HPC in their experiments. Even though ROOT will still be the preferred choice for the custodial data-sets for the analysis, HPC-friendly I/O libraries like HDF5 maybe beneficial for the data-processing.
 HEP community needs to start investigating the data storage formats that are optimized for the HPC environment in terms of HEP requirements as discussed in this paper. At the same time, the HEP experiments can try to identify the areas where HPC-friendly libraries such as HDF5 can be adopted with or without utilizing ROOT-serialization. In the long run, data generated from intensive computation should be designed to be directly coupled with these HPC-friendly formats. The computing requirements of the future HEP experiments can be fulfilled by increasing the share of HPC in all the relevant areas. Fully utilizing the HPC performance will require adopting the HPC-friendly data storage systems.\\
 Exploring the HPC resources will require new or increased efforts from the HEP community to identify the areas where the HPC resources can be adopted without compromising the gains made in the last two decades. At the same time, HEP community should increase the engagement with the HPC experts to integrate the HPC related tools and technology into the HEP work-flow. The HEP experiments will see increased role of HPC and the exploratory works (as discussed in this paper) will help the HEP community to adopt these new resources and sustain the computational requirements well into the future.
\newpage

\bibliographystyle{unsrt}
\bibliography{main}

\end{document}